\documentstyle[aps,prl,preprint,floats,epsfig]{revtex}

\newcommand{\aerr}[3]   {\mbox{${{#1}^{+ #2}_{- #3}}$}}

\newcommand{\KS}{\mbox{$K_S^0$}}

\begin{document}
\draft    
                        
\preprint{\tighten\vbox{\hbox{\hfil CLNS 99/1650} 
                        \hbox{\hfil CLEO 99-18}   
}}

\title{Study of Two-Body $B$ Decays to Kaons and Pions:
Observation of $B\to\pi^+\pi^-$, $B\to K^{\pm}\pi^0$, and $B\to K^0\pi^0$ 
Decays.} 
 
\author{(CLEO Collaboration)}
\date{\today}
\maketitle
\tighten

\begin{abstract} 
We have studied charmless hadronic decays of $B$ mesons into two-body
final states with kaons and pions and observe three new processes with 
the following branching fractions:
${\cal{B}}(B \rightarrow \pi^+\pi^-) = (4.3^{+1.6}_{-1.4} \pm 0.5) \times 10^{-6}$, 
${\cal{B}}(B \rightarrow K^0 \pi^0) = (14.6^{+5.9+2.4}_{-5.1-3.3}) \times
10^{-6}$, and
${\cal{B}}(B \rightarrow K^{\pm} \pi^0) = (11.6^{+3.0+1.4}_{-2.7-1.3}) \times
10^{-6}$.
 We also update our previous measurements for the decays $B\to
K^{\pm}\pi^{\mp}$ and $B^{\pm}\to K^0\pi^{\pm}$.
\end{abstract}
\newpage

\renewcommand{\thefootnote}{\fnsymbol{footnote}}

\begin{center}
D.~Cronin-Hennessy,$^{1}$ Y.~Kwon,$^{1,}$%
\footnote{Permanent address: Yonsei University, Seoul 120-749, Korea.}
A.L.~Lyon,$^{1}$ E.~H.~Thorndike,$^{1}$
C.~P.~Jessop,$^{2}$ H.~Marsiske,$^{2}$ M.~L.~Perl,$^{2}$
V.~Savinov,$^{2}$ D.~Ugolini,$^{2}$ X.~Zhou,$^{2}$
% ...
\end{center}
 
\small
\begin{center}
$^{1}${University of Rochester, Rochester, New York 14627}\\
$^{2}${Stanford Linear Accelerator Center, Stanford University, Stanford,
California 94309}\\
% ...
\end{center}
\setcounter{footnote}{0}
\newpage

%
%
  CP violation in the Standard Model (SM) arises naturally from the complex
phase in the Cabibbo-Kobayashi-Maskawa (CKM) quark-mixing matrix\cite{CKM}. 
This picture is supported by numerous experimental
constraints\cite{review},
as well as recent observation of direct CP violation in the 
kaon system\cite{ktev}, but it remains an open
experimental question whether this phase is the only source of CP 
violation in nature. Studies of the rare charmless decays of $B$ mesons
are likely to play an important role in constraining the CKM matrix and
testing the SM picture of CP violation.

Several approaches have been suggested to extract this phase
information from measurements of rare $B$ decays.
 Ratios of various $B\to K\pi$ branching fractions were
shown\cite{neubert-rosner} to depend explicitly on
 $\gamma \equiv Arg(V^*_{ub})$ 
with relatively modest model dependence.
 Within a factorization model,
branching fractions of a large number of rare
$B$ decays can be parametrized
by a small number of independent
physical quantities, including $\gamma$, which can then be
extracted through a global fit\cite{hsw} to existing data.
 Finally, measurement of the time-dependent CP-violating asymmetry in  
the decay $B^0\to\pi^+\pi^-$ can be
used to determine the sum of $\gamma$ and
the phase $\beta \equiv Arg(V^*_{td})$.  In this case additional
measurements of other isospin-related $B\to \pi\pi$ processes are
required to allow extraction of $\gamma + \beta$ \cite{isospin}.

In this Letter we present new measurements of $B\to K\pi$ and
$B\to\pi\pi$ branching fractions with significantly 
increased statistics,
superseding results from our previous publication\cite{prl98}.
In particular we present first observations of the
long-awaited mode $B\to\pi^+\pi^-$, as well as $B\to K^{\pm}\pi^0$ and
 $B\to K^0\pi^0$ decays.

%
%
The data used in this analysis were collected with the CLEO II 
detector at the Cornell Electron Storage Ring (CESR).  It
consists of $9.13~{\rm fb}^{-1}$ taken at the $\Upsilon$(4S),
corresponding to 9.66M $B\bar{B}$ pairs, and $4.35~{\rm
fb}^{-1}$ taken below $B\bar{B}$ threshold, used for continuum
background studies.

CLEO II is a general purpose solenoidal magnet detector,
described in detail elsewhere~\cite{detector}.  Cylindrical drift
chambers in a 1.5T solenoidal magnetic field measure momentum and
specific ionization ($dE/dx$) of charged particles. Photons are detected
using a 7800-crystal CsI(Tl) electromagnetic calorimeter.  In the CLEO II.V
detector configuration, the innermost chamber was replaced by a
 3-layer, double-sided silicon
vertex detector, and the gas in the main drift chamber was changed
from an argon-ethane to a helium-propane mixture. As a result of these 
modifications, the CLEO II.V portion of the data (2/3 of the total) has
significantly improved particle identification and momentum resolution.

Efficient track quality requirements are imposed on charged tracks,
and pions and kaons are identified by $dE/dx$.  The separation between
kaons and pions for typical signal momenta $p \sim 2.6$~GeV$/c$\ is
$1.7$ standard deviations ($\sigma$) for CLEO II data and
$2.0~\sigma$ for
CLEO II.V data.  Candidate \KS\ are selected from pairs of tracks
forming well-measured displaced vertices with a $\pi^+\pi^-$
invariant mass within $2\sigma$ of the nominal \KS\ mass. Pairs of
photons with an invariant mass within 2.5$\sigma$ of the nominal
$\pi^0$ mass are kinematically fitted with the mass
constrained to the nominal $\pi^0$ mass.
  Electrons are rejected based on $dE/dx$ and the ratio of
the track momentum to the associated shower energy in the CsI
calorimeter; muons are rejected based on the penetration depth in
the instrumented steel flux return.  

%
%
  The $B$ decay candidate is identified via invariant mass and total energy
of its decay products. 
  We calculate a beam-constrained $B$ mass
$M = \sqrt{E_{\rm b}^2 - p_B^2}$, where $p_B$ is the $B$ candidate
momentum and $E_{\rm b}$ is the beam energy.  The resolution in $M$\
is dominated by the beam energy spread and ranges from 2.5 to
3.0~${\rm MeV}$, where the larger resolution corresponds to
decay modes with a $\pi^0$.  We define $\Delta E = E_1 + E_2 - E_{\rm
b}$, where $E_1$ and $E_2$ are the energies of the daughters of the
$B$ meson candidate.  The resolution in $\Delta E$ is mode-dependent.
For final states without $\pi^0$'s, the $\Delta E$ resolution is
$20$~MeV ($25$ MeV in CLEO II). For modes with $\pi^0$'s
the $\Delta E$ resolution is worse by approximately a factor
of two and becomes slightly asymmetric because of energy loss out of
the back of the CsI crystals.  
 We accept events with $M$ within
$5.2-5.3$~$\rm {GeV}$\ and $|\Delta E|<200$~MeV. This fiducial
region includes the signal region and a generous sideband for
background normalization.
$\pi\pi$ and $K\pi$ signal events are distinguished both by $dE/dx$ and
$\Delta E$ observables. The $\Delta E$ distribution for
 $B \rightarrow K^+ \pi^-$,
calculated under the replacement of $m_K$ by $m_\pi$, is centered at
-42~MeV, giving a separation of $2.1\sigma$($1.7\sigma$ in CLEO II)
between $B \rightarrow K^+ \pi^-$ and $B \rightarrow \pi^+\pi^-$. 

We have studied backgrounds from $b\to c$\ decays and other $b\to u$\
and $b\to s$\ decays and find that all are negligible for the analyses
presented here. The main background arises from $e^+e^-\to q\bar q$\
(where $q=u,d,s,c$).  Such events typically exhibit a two-jet
structure and can produce high momentum back-to-back tracks in the
fiducial region.  To reduce contamination from these events, we
calculate the angle $\theta_S$ between the sphericity axis\cite{shapes}
of the candidate tracks and showers and the sphericity axis of the
rest of the event. The distribution of $\cos\theta_S$\ is strongly
peaked at $\pm 1$ for $q\bar q$\ events and is nearly flat for $B\bar
B$\ events. We require $|\cos\theta_S|<0.8$\ which eliminates $83\%$\
of the background.  Using a detailed GEANT-based Monte Carlo
simulation\cite{geant} we determine overall detection efficiencies
${\cal E}$ of $11-48\%$, as listed in
Table~\ref{tab:results}. Efficiencies include the branching fractions
for $K^0\to K^0_S\to \pi^+\pi^-$\ and $\pi^0\to \gamma\gamma$ where
applicable.

Additional discrimination between isotropic signal and rather jetty 
$q\bar q$\ background is provided by the
cosine of the angle between the candidate sphericity axis and beam
axis (expected to be isotropic for signal,  $1+cos^2\theta$
distribution for $q\bar{q}$ 
background); the ratio of Fox-Wolfram moments $H_2/H_0$~\cite{fox}
(expected to be smaller for signal than for background); and the
distribution of the energy from the rest of the event relative to the
candidate's sphericity axis, as characterized by the energy in nine 
$10^\circ$ angular bins. These 11 variables are combined by a Fisher 
discriminant technique as described in detail in
Ref.~\cite{bigrare}  The Fisher discriminant is a linear combination
of experimental observables
${\cal F}\equiv \sum_{i=1}^{N}\alpha_i y_i$,\ where the coefficients
$\alpha_i$ are chosen to maximize the separation between the simulated 
signal and background samples.

We perform unbinned
maximum-likelihood fits using $\Delta E$, $M$, ${\cal F}$,
the angle between the $B$ meson momentum and beam axis, and $dE/dx$
(where applicable) as input information for each candidate event to
determine the signal yields.  Four different fits are performed, one
for each topology ($h^+h^-$, $h^\pm \pi^0$, $h^\pm K^0_S $, and
$K^0_S\pi^0$, $h^\pm$\ referring to a charged kaon or pion).  In each
of these fits, the likelihood of the event is parametrized by the sum
of probabilities for all relevant signal and background hypotheses,
with relative weights determined by maximizing the likelihood function
$\cal L$.  The probability of a particular hypothesis is calculated
as a product of the probability density functions (PDFs) for each of
the input variables.  Further details about the likelihood fit can be
found in Ref.~\cite{bigrare}.  The parameters for the PDFs are
determined from independent data and high-statistics Monte Carlo
samples. We estimate a systematic error on the fitted yield by varying
the PDFs used in the fit within their uncertainties.  These
uncertainties are dominated by the limited statistics in the
independent data samples we used to determine the PDFs.  The
systematic errors on the measured branching fractions are obtained by
adding this fit systematic in quadrature with the systematic error on
the efficiency.

%
%

\begin{table}
\begin{center}
\caption{Summary of experimental results.  The errors on
branching fractions ${\cal B}$ are statistical and systematic
respectively. Reconstruction efficiency ${\cal E}$ includes branching
fractions of $K^0 \to K^0_S \to \pi^+\pi^-$ and $\pi^0\to
\gamma\gamma$ when applicable. 
We assume equal branching fraction for $\Upsilon(4s)\to B^0\bar B^0$
and $B^+B^-$.
Theoretical predictions are taken from Ref. 13.} 
\begin {tabular}{l c c c c c}
Mode&   $N_S$ & Sig. & ${\cal E}(\%)$ &${\cal B}\times 10^{6}$ & Theory ${\cal B}\times 10^{6}$\\
\hline
$\pi^+\pi^-$ &
$20.0^{+7.6}_{-6.5}$  & 
4.2$\sigma$        &
$48$ & 
$\aerr{4.3}{1.6}{1.4}\pm0.5$  & 
8--26      \\
$\pi^\pm\pi^0$ & 
$21.3^{+9.7}_{-8.5}$ & 
3.2$\sigma$        &
$39$ & 
$<12.7$  & 
3--20        \\
%
%
%
%
%
%
\hline
$K^\pm\pi^\mp$   &
$80.2^{+11.8}_{-11.0}$ & 
11.7$\sigma$         &
$48$ & 
$17.2^{+2.5}_{-2.4}\pm 1.2$ & 
7--24        \\
$K^\pm\pi^0$   &
$42.1^{+10.9}_{-9.9}$  & 
6.1$\sigma$        &
$38$ & 
$\aerr{11.6}{3.0}{2.7}\aerr{}{1.4}{1.3}$  & 
3--15         \\
$K^0\pi^\pm$   &
$25.2^{+6.4}_{-5.6}$  & 
7.6$\sigma$       &
$14$
&$18.2^{+4.6}_{-4.0}\pm1.6$ &  
8--26               \\
$K^0\pi^0$   & 
$16.1^{+5.9}_{-5.0}$ & 
4.9$\sigma$        &
$11$
& $\aerr{14.6}{5.9}{5.1}\aerr{}{2.4}{3.3}$  & 
3--9     \\
\hline
$K^+ K^-$     &
$0.7^{+3.4}_{-0.7}$   & 
0.0$\sigma$       &
$48$ 
& $<1.9$ & 
      --              \\
$K^{\pm}\bar{K}^0$   & 
$1.4^{+2.4}_{-1.3}$  & 
1.1$\sigma$       &
$14$
& $<5.1$  & 
0.7--1.5   \\
%
%
%
%
%
\end {tabular}
\label{tab:results}
\end{center}
\end {table}

\begin{figure}[hbp]
\centering
\leavevmode
\epsfxsize=3.25in
\epsffile{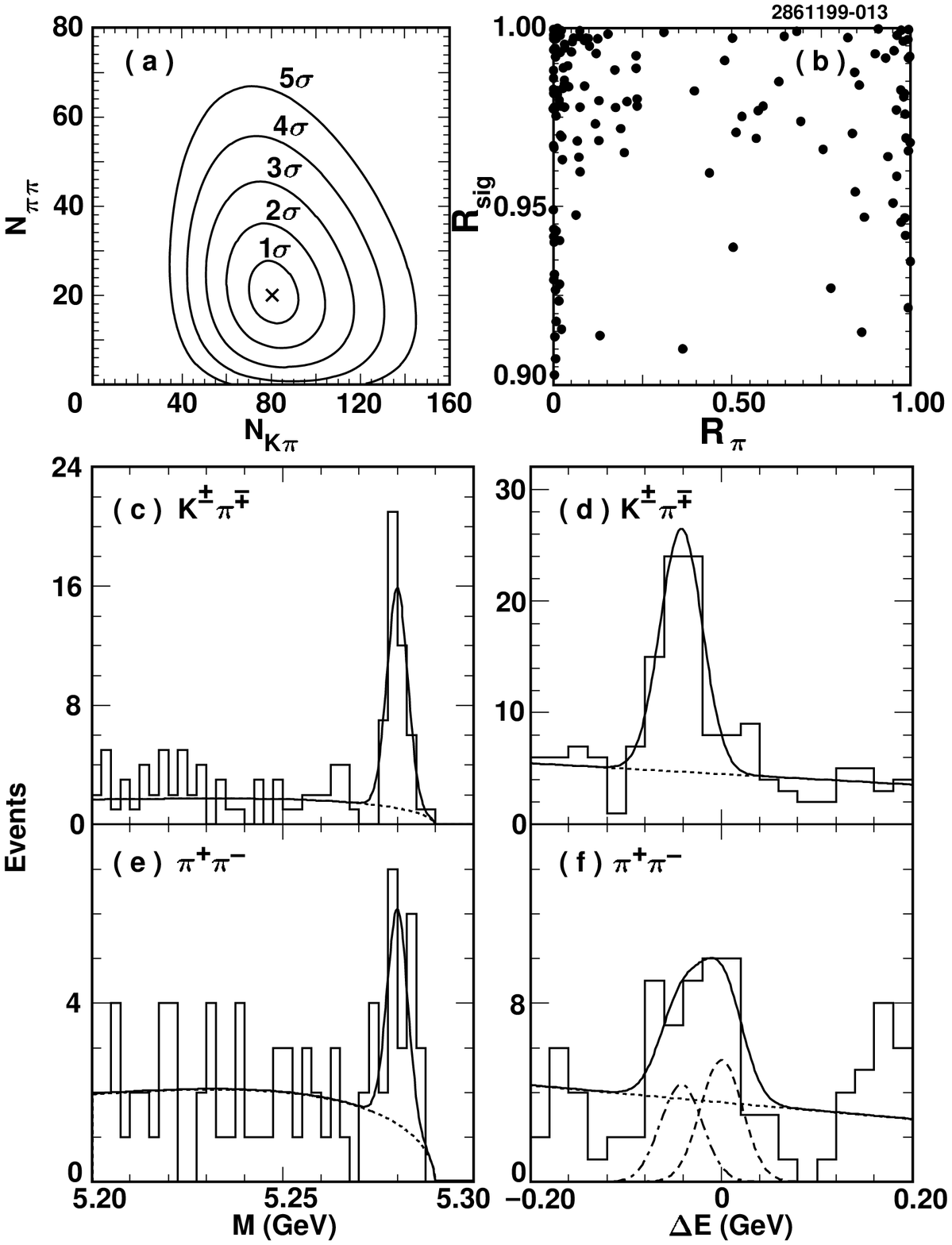}
\caption{
Results for $B\to K^\pm\pi^\mp$ and $B\to \pi^+\pi^-$.
 Contours of the likelihood function versus $K\pi$ 
and $\pi\pi$ event yield (a);
 likelihood ratios (b) $-$ signal events cluster near the top of the figure,
 and separate into
$K\pi$-like events on the left and $\pi\pi$-like events on the right;
 beam constrained mass for $K\pi$-like events (c);
 $\Delta E$ for $K\pi$-like events (d);
 beam constrained mass for $\pi\pi$-like events (e);
 $\Delta E$ for $\pi\pi$-like events (f), with both $\pi\pi$ signal (dashed
line) and $K\pi$ cross-feed (dot-dashed line) shown.}
\label{fig:one}
\end{figure}

Figure \ref{fig:one}a shows the results of the likelihood fit for $B\to
\pi^+\pi^-$ and $B\to K^\pm\pi^\mp$.  The curves represent the
$n\sigma$ contours, which correspond to the increase in $-2\ln{\cal
L}$ by $n^2$. Systematic uncertainties are not included in any contour
plots.  The statistical significance of a given signal yield is
determined by repeating the fit with the signal yield fixed to be zero
and recording the change in $-2\ln{\cal L}$.
We also compute from the PDFs the event-by-event probability to be
signal or continuum background, as well as the probability to be
$K\pi$-like or $\pi\pi$-like. From these we form likelihood ratios,
${\cal R}_{sig} = (P^s_{\pi\pi}+P^s_{K\pi})/
(P^s_{\pi\pi}+P^s_{K\pi}+P^c_{\pi\pi}+P^c_{K\pi}+P^c_{KK})$ and ${\cal
R}_{\pi} = P^s_{\pi\pi}/(P^s_{\pi\pi}+P^s_{K\pi})$.  Superscripts $s$
and $c$ denote signal and continuum background respectively.
Figure \ref{fig:one}b illustrates the distribution of events in
${\cal R}_{sig}$ (vertical axis) and ${\cal R}_\pi$ (horizontal axis).
The cluster of events in the upper right corner is clear evidence for
$B\to \pi^+\pi^-$.

\begin{figure}[hbp]
\centering
\leavevmode
\epsfxsize=3.25in
\epsffile{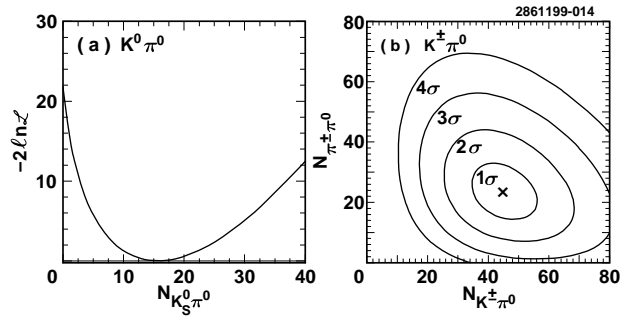}
\caption{
Likelihood function versus event yield for $B\to K^0\pi^0$ (a) 
and likelihood contours for $B\to h^{\pm}\pi^0$ (b).}
\label{fig:two}
\end{figure}

\begin{figure}[hbp]
\centering
\leavevmode
\epsfxsize=3.25in
\epsffile{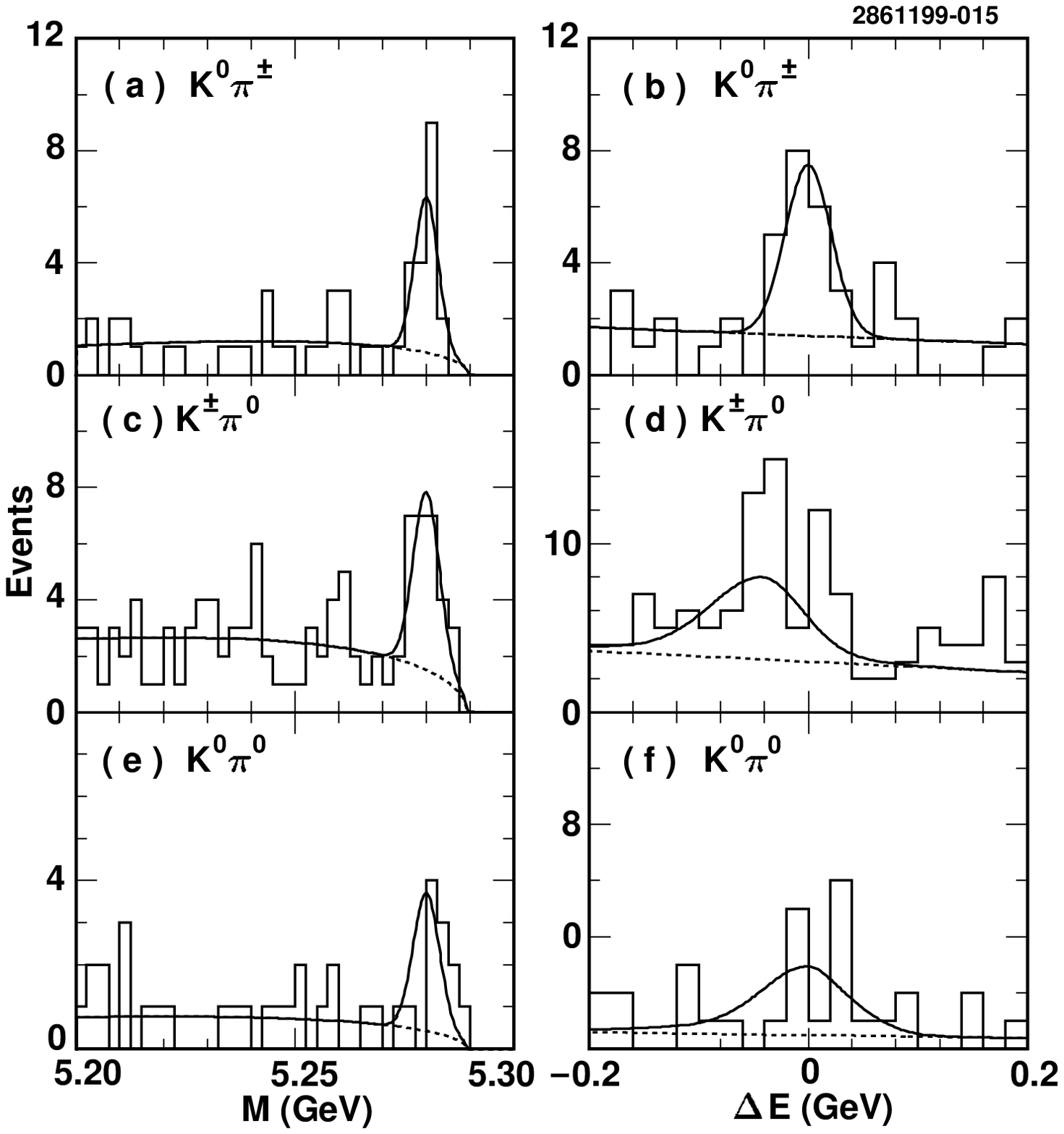}
\caption{
Beam constrained mass and $\Delta E$ distributions for 
 $B\to K^0\pi^{\pm}$ (a-b),
 $B\to K^{\pm}\pi^0$ (c-d),
 and $B\to K^0\pi^0$ (e-f).}
\label{fig:three}
\end{figure}

Figures \ref{fig:one}(c-f) show distributions in $M$ and
$\Delta E$ for events after cuts on likelihood ratios ${\cal R}_{sig}$ and 
${\cal R}_\pi$ computed without $M$ and
$\Delta E$, respectively.
The likelihood fit projections for signal and 
background components,
suitably scaled to account for the
efficiencies of the additional cuts (50-70 \% for signal), are overlaid.   
Figure \ref{fig:two} shows the likelihood functions for the fits
to $B\to K^0\pi^0$ and $B\to h^{\pm}\pi^0$.
Figure \ref{fig:three} shows $M$ and $\Delta E$ distributions for
 $B\to K^0\pi^\pm$,
 $B\to K^\pm\pi^0$,
 and $B\to K^0\pi^0$. 

We summarize all branching fractions and upper limits in Table
\ref{tab:results}.  In addition to the first observations
$B\to\pi^+\pi^-$, $B \to K^+ \pi^0$, and $B \to K^0 \pi^0$, we report
improved measurements for the decays $B \to
K^\pm \pi^\mp$ and $B \to K^0 \pi^\pm$.
The table also includes
a range of theoretical predictions taken from recent literature\cite{all}.
We see some indication for the decay $B\to \pi^{\pm}\pi^0$ with the branching 
fraction of 
${\cal B}(B\to \pi^{\pm}\pi^0)=(5.6^{+2.6}_{-2.3} \pm 1.7)\times 10^{-6}$,
but statistical significance of the signal yield is insufficient to claim an
observation for this decay mode. 
We find no evidence for the decays $B\to K^+ K^-$ and $B\to K^\pm
K^0$,
and calculate $90\%$\ confidence level (CL) upper limit 
yields by integrating the likelihood function
\begin{equation}
{\int_0^{N^{UL}} {\cal L}_{\rm max} (N) dN
\over
\int_0^{\infty} {\cal L}_{\rm max} (N) dN}
= 0.90
\nonumber
\end{equation}
where ${\cal L}_{\rm max}(N)$ is the maximum $\cal L$\ at fixed $N$\
to conservatively account for possible correlations among the free
parameters in the fit. We then increase upper limit yields by their
systematic errors and reduce detection efficiencies by their
systematic errors to calculate branching fraction upper limits given
in Table I. 

To evaluate how systematic uncertainties in the PDFs affect the
statistical significance for modes where we report first observations,
we repeated the fits for the $h^+h^-,\ h^+\pi^0$ and $ K^0\pi^0$ modes 
with all
PDFs changed simultaneously within their uncertainties to maximally
reduce the signal yield in the modes of interest. Under these extreme
conditions, the significance of the first-observation modes $\pi^+
\pi^-,\ K^+\pi^0$ and $K^0 \pi^0$ becomes $3.2,\ 5.3$ and $3.8~\sigma$
respectively. We also evaluate the branching fractions with alternative
analyses using tighter and looser cuts on the continuum-suppressing 
variable $|\cos\theta_S|$.  These variations correspond
to halving and doubling the background in the fitted sample.  The
changes in branching fractions under these variations are insignificant
compared to the statistical error of our results.

The ratio of the branching fractions 
${\cal B}(B\to K^{\pm}K^0)/{\cal B}(B\to \pi^{\pm}K^0)$ can be used
to estimate the size of final state interactions in charmless rare 
$B$ decays\cite{petrov}. Following the method outlined above we 
calculate ${\cal B}(B\to K^{\pm}K^0)/{\cal B}(B\to \pi^{\pm}K^0) < 0.3$ 
at 90\% CL. 
It has also been suggested\cite{jerome} to use the ratio of the branching
fractions 
${\cal B}(B\to K^{\pm}\pi^{\mp})/{\cal B}(B\to \pi^+\pi^-)$ to estimate
uncertainties in the measurement of the unitarity triangle
parameter $\alpha = \pi - \beta - \gamma$ via 
$B^0(t)\to \pi^+\pi^-$. We obtain 
${\cal B}(B\to K^\pm\pi^\mp)/{\cal B}(B\to \pi^+\pi^-) < 15 $
at 90\% CL which implies that an error on $\alpha$ obtained from time-dependent
asymmetry measurements of $B^0 \to \pi^+\pi^-$ can be as high as $60^\circ$ \cite{jerome}.   

%
%
In summary: we have made a first observation of $B\to \pi^+\pi^-$;
measured branching fractions for all four
exclusive $B\to K\pi$, including first observations of the decays
$B\to K^+\pi^0$ and $B\to K^0\pi^0$; 
obtained improved upper limits on 
$B\to \pi^+\pi^0$ and $B\to K\bar{K}$ modes. The hierarchy of branching 
fractions $ KK < \pi\pi < K\pi$ is obvious.

%
%
We thank W.-S. Hou for many useful discussions.
We gratefully acknowledge the effort of the CESR staff in providing us
with excellent luminosity and running conditions. This work was
supported
 by the National Science
Foundation, the U.S. Department of Energy, the Research Corporation, 
the Natural Sciences
and Engineering Research Council of Canada, the A.P. Sloan Foundation,
the Swiss National Science Foundation, and Alexander von Humboldt Stiftung.



\begin{thebibliography}{99}


\bibitem{CKM} 
M.~Kobayashi and K.~Maskawa, Prog.\ Theor.\ Phys.\ {\bf 49}, 652 (1973).


\bibitem{review} 
A.J. Buras, Lectures given at the 14th Lake Louise Winter
Institute, February 1999,  hep-ph/9905437.

\bibitem{ktev}
A.~Alavi-Harati {\it et al.} (KTeV), Phys. Rev. Lett. {\bf 83}, 22 (1999);
V.~Fanti {\it et al.} (NA48). Phys. Lett. B {\bf 465}, 335 (1999).


\bibitem{neubert-rosner} 
M.~Gronau, J.~L.~Rosner, and D.~London, 
Phys.\ Rev.\ Lett.\ {\bf 73}, 21 (1994);
R.~Fleischer, 
Phys.\ Lett.\ B {\bf 365}, 399 (1996);
R.~Fleischer and T.~Mannel,
Phys.\ Rev.\ D {\bf 57}, 2752 (1998).
M. Neubert and J. Rosner,
Phys. Lett. B {\bf 441}, 403 (1998);
M. Neubert, 
J. High Energy Phys. {\bf 02} (1999) 014.


\bibitem{hsw}
W.-S.~Hou, J.~G.~Smith, and F.~W\"urthwein, hep-ex/9910014.




\bibitem{isospin} 
M.~Gronau and D.~London,  Phys.\ Rev.\ Lett.\ {\bf 65}, 3381 (1990). 
 

\bibitem{prl98}
\label{refprl98}
R.~Godang {\it et al.} (CLEO Collaboration),
Phys.\ Rev.\ Lett. {\bf 80}, 3456 (1998).



\bibitem{detector}
Y.~Kubota {\it et al.} (CLEO Collaboration),
Nucl.\ Instrum.\ Methods Phys.\ Res., Sec.\ A{\bf 320}, 66 (1992);
T.S.~Hill, Nucl.\ Instrum.\ Methods Phys.\ Res., Sec.\ A {\bf 418},
32 (1998).


\bibitem{shapes} S.~L.~Wu, Phys.\ Rep.\ C{\bf 107}, 59 (1984).

\bibitem{geant} R.~Brun {\it et al.}, GEANT 3.15, CERN DD/EE/84-1.


\bibitem{fox} G.~Fox and S.~Wolfram, Phys.\ Rev.\ Lett.\ {\bf 41}, 
1581 (1978).
\bibitem{bigrare}
D.~M.~Asner {\it et al.} (CLEO Collaboration),
Phys.\ Rev.\ D {\bf 53}, 1039 (1996).



\bibitem{all}
\label{refall}
N.~G.~ Deshpande and J.~ Trampetic, Phys.\ Rev.\ D {\bf 41}, 895 (1990);
L.-L.~Chau {\it et al.}, Phys.\ Rev.\ D {\bf 43}, 2176 (1991);
A.~ Deandrea {\it et al.}, Phys.\ Lett.\ B {\bf 318}, 549 (1993);
A.~ Deandrea {\it et al.}, Phys.\ Lett.\ B {\bf 320}, 170 (1994);
G.~Kramer, W.~F.~Palmer, and H.~Simma, Z.\ Phys.\ C {\bf 66}, 429 (1995);
G.~Kramer and W.~F.~Palmer, Phys.\ Rev.\ D {\bf 52}, 6411 (1995);
D.~Ebert, R.~N.~Faustov, and V.~O.~Galkin, Phys.\ Rev.\ D {\bf 56}, 312 (1997);
D.~Du and L.~Guo, Zeit.\ Phys.\ C {\bf 75}, 9 (1997);
N.G.~Deshpande, B.~Dutta, and S.~Oh, Phys.\ Rev.\ D {\bf 57}, 5723 (1998);
H.~Cheng and B.~Tseng, Phys.\ Rev.\ D {\bf 58}, (1998) 094005;
A.~Ali, G.~Kramer, and C.~L\"u, Phys.\ Rev.\ D {\bf 58}, (1998) 094009.

\bibitem{petrov}
A.~F.~Falk, A.~L.~Kagan, Y.~Nir, and A.~A.~Petrov, 
Phys.\ Rev.\ D {\bf 57}, 4290 (1998).

\bibitem{jerome}
J.~Charles, Phys.\ Rev.\ D {\bf 59}, 054007 (1999).

\end{thebibliography}
\end{document}